# SKIP SEQUENCING: A DECISION PROBLEM IN QUESTIONNAIRE DESIGN

By Charles F. Manski[1] and Francesca Molinari[2]

*Northwestern University and Cornell University*

This paper studies questionnaire design as a formal decision problem, focusing on one element of the design process: skip sequencing. We propose that a survey planner use an explicit loss function to quantify the trade-off between cost and informativeness of the survey and aim to make a design choice that minimizes loss. We pose a choice between three options: ask all respondents about an item of interest, use skip sequencing, thereby asking the item only of respondents who give a certain answer to an opening question, or do not ask the item at all. The first option is most informative but also most costly. The use of skip sequencing reduces respondent burden and the cost of interviewing, but may spread data quality problems across survey items, thereby reducing informativeness. The last option has no cost but is completely uninformative about the item of interest. We show how the planner may choose among these three options in the presence of two inferential problems, item nonresponse and response error.

**1. Introduction.** Designing a questionnaire for administration to a sample of respondents requires many decisions about the items to be asked, the wording and ordering of the questions, and so on. Considerable research has investigated the item response rates and patterns associated with alternative designs. See Krosnick (1999) for a recent review of the literature. Researchers have also called attention to the tension between the desire to reduce the costs and increase the informativeness of surveys. See, for example, Groves (1987) and Groves and Heeringa (2006). However, survey researchers have not studied questionnaire design as a formal decision problem in which one

Received July 2007; revised September 2007.
[1]Supported in part by National Institute of Aging Grants R21 AG028465-01 and 5P01AG026571-02, and by NSF Grant SES-05-49544.
[2]Supported in part by National Institute of Aging Grant R21 AG028465-01 and by NSF Grant SES-06-17482.

uses an explicit loss function to quantify the trade-off between cost and informativeness and aims to make a design choice that minimizes loss. This paper takes an initial step in that direction. We consider one element of the design problem, the use of skip sequencing.

*Skip sequencing* is a widespread survey practice in which the response to an opening question is used to determine whether a respondent should be asked certain subsequent questions. The objective is to eliminate inapplicable questions, thereby reducing respondent burden and the cost of interviewing. However, skip sequencing can amplify data quality problems. In particular, skip sequencing exacerbates the identification problems caused by item nonresponse and response errors.

A respondent may not answer the opening question. When this happens, a common practice is to label the subsequent questions as inapplicable. However, they may be applicable, in which case the item nonresponse problem is amplified. Another practice is to impute the answer to the opening question and, if the imputation is positive, to also impute answers to the subsequent questions. Some of these imputations will inevitably be incorrect. A particularly odd situation occurs when the answer to the opening question should be negative but the imputation is positive. Then answers are imputed to subsequent questions that actually are inapplicable.

A respondent may answer the opening question with error. An error may cause subsequent questions to be skipped, when they should be asked, or vice versa. An error of the first type induces nonresponse to the subsequent questions. The consequences of an error of the second type depend on how the respondent answers the subsequent questions, having answered the opening one incorrectly.

ILLUSTRATION 1. The 2006 wave of the Health and Retirement Study (HRS) asked current Social Security recipients about their expectations for the future of the Social Security system. An opening question asked broadly: "Thinking of the Social Security program in general and not just your own Social Security benefits: On a scale from 0 to 100 (where 0 means no chance and 100 means absolutely certain), what is the percent chance that Congress will change Social Security sometime in the next 10 years, so that it becomes less generous than now?" If the answer was a number greater than zero, a follow-up question asked "We just asked you about changes to Social Security in general. Now we would like to know whether you think these Social Security changes might affect your own benefits. On a scale from 0 to 100, what do you think is the percent chance that the benefits you yourself are receiving from Social Security will be cut some time over the next 10 years?" If a person did not respond to the opening question or gave an answer of 0, the follow-up question was not asked.

ILLUSTRATION 2. The 1990 wave of the National Longitudinal Survey of Older Men (NLSOM) queried respondents about their limitations in activities of daily living (ADLs). An opening question asked broadly: "Because of a health or physical problem, do you ever need help from anyone in looking after personal care such as dressing, bathing, eating, going to the bathroom, or other such daily activities?" If the answer was positive, the respondent was then asked if he/she receives help from another person in each of six specific ADLs (bathing/showering, dressing, eating, getting in or out of a chair or bed, walking, using the toilet). If the answer was negative or missing, the subsequent questions were skipped out.

These illustrative uses of skip sequencing save survey costs by asking a broad question first and by following up with a more specific question only when the answer to the broad question meets specified criteria. However, nonresponse or response error to the opening question may compromise the quality of the data obtained.

This paper studies skip sequencing as a decision problem in questionnaire design. We suppose that a survey planner is considering whether and how to ask about an item of interest. Three design options follow:

> Option All ($A$): ask all respondents the question.
> Option Skip ($S$): ask only those respondents who respond positively to an opening question.
> Option None ($N$): do not ask the question at all.

These options vary in the cost of administering the questions and in the informativeness of the data they yield. Option ($A$) is most costly and is potentially most informative. Option ($S$) is less costly but may be less informative if the opening question has nonresponse or response errors. Option ($N$) has no cost but is uninformative about the item of interest. We suppose that the planner must choose among these options, weighing cost and informativeness as he deems appropriate. We suggest an approach to this decision problem and give illustrative applications.

The paper is organized as follows. As a prelude, Section 2 summarizes the few precedent studies that consider the data quality aspects of skip sequencing. These studies do not analyze skip sequencing as a decision problem.

Section 3 formalizes the problem of choice among design options. We assume that the survey planner wants to minimize a loss function whose value depends on the cost of a design option and its informativeness. Thus, evaluation of the design options requires that the planner measure their cost and informativeness.

Suppose that a planner wants to combine sample data on an item with specified assumptions in order to learn about a population parameter of interest. When the sample size is large, we propose that informativeness be measured by the size of the identification region that a design option yields for this parameter. As explained in Manski (2003), the *identification region* for the parameter is the set of values that remain feasible when unlimited observations from the sampling process are combined with the maintained assumptions. The parameter is *point-identified* when this set contains a single value and is *partially identified* when the set is smaller than the parameter's logical range, but is not a single point. In survey settings with large samples of respondents, where identification rather than statistical inference is the dominant inferential problem, we think it natural to measure informativeness by the size of the identification region. The smaller the identification region, the better. Section 6 discusses measurement of informativeness when the sample size is small. Then confidence intervals for the partially identified parameter may be used to measure informativeness.

Sections 4 and 5 apply the general ideas of Section 3 in two polar settings having distinct inferential problems. Section 4 studies cases in which there may be nonresponse to the questions posed but it is assumed that there are no response errors. We first derive the identification regions under options $A$, $S$ and $N$. We then show the circumstances in which a survey planner should choose each option. To illustrate, we consider choice among options for querying respondents about their expectations for future personal Social Security benefits. The HRS 2006 used skip sequencing, as described in Illustration 1. Another option would be to ask all respondents both the broad and the personal question. A third option would be to ask only the broad question, omitting the one about future personal benefits.

Section 5 studies the other polar setting in which there is full response but there may be response errors. Again, we first derive the identification regions under the three design options and then show when a survey planner should choose each option. To illustrate, we consider choice among options for querying respondents about limitations in ADLs. The NLSOM used skip sequencing, as described in Illustration 2. Another survey, the 1993 wave of the Assets and Health Dynamics Among the Oldest Old (AHEAD) asked all respondents about a set of specific ADLs. A third option would be to not ask about specific ADLs at all.

Section 6 concludes by calling for further analysis of questionnaire design as a decision problem.

**2. Previous studies of skip sequencing.** As far as we are aware, there has been no precedent research studying skip sequencing as a decision problem in questionnaire design. Messmer and Seymour (1982) and Hill (1991, 1993) are the only precedent studies recognizing that skip sequencing may amplify data quality problems.

Messmer and Seymour studied the effect of skip sequencing on item nonresponse in a large scale mail survey. Their analysis asked whether the difficult structure of the survey, particularly the fact that respondents were instructed to skip to other questions perhaps several pages away in the questionnaire, increased the number of unanswered questions. Their analysis indicates that branching instructions significantly increased the rate of item nonresponse for questions following a branch, and that this effect was higher for older individuals. This work is interesting but it does not have direct implications for modern surveys, where skip sequencing is automated rather than performed manually.

Hill used data from five interview/reinterview sequence pairs in the 1984 Survey of Income and Program Participation (SIPP) Reinterview Program. He examined data errors that manifest themselves through a discrepancy between the responses given in the two interviews, and categorized these discrepancies in three groups. In his terminology, a response discrepancy occurs when a different answer is recorded for an opening question in the interview and in the reinterview. A response induced sequencing discrepancy occurs when, as a consequence of different answers to the opening question, a subsequent question is asked in only one of the two interviews. A procedurally induced sequencing discrepancy occurs when, in one of the two interviews but not both, an opening question is not asked and, therefore, the subsequent question is not asked either.

Hill used a discrete contagious regression model to assess the relative importance of these errors in reducing data quality. The contagion process was used to express the idea that error spreads from one question to the next via skip sequencing. Within this model, the "conditional population at risk of contagion" expresses the idea that the number of remaining questions in the sequence at the point where the initiating error occurs gives an upper bound on the number of errors that can be induced. Hill's results suggest that the losses of data reliability caused by induced sequencing errors are at least as large as those induced by response errors. Moreover, the relative importance of sequencing errors strongly increases with the sequence length. This suggests that the reliability of individual items will be lower, all else equal, the later they appear in the sequence.

**3. A formal design problem.**

3.1. *The choice setting.* We pose here a formal questionnaire design problem that highlights how skip sequencing may affect data quality. To focus on this matter, we find it helpful to simplify the choice setting in three major respects.

First, we suppose that a large random sample of respondents is drawn from a much larger population. This brings identification to the fore as the dominant inferential problem, the statistical precision of sample estimates receding into the background as a minor concern. We also suppose that all sample members agree to be interviewed. Hence, inferential problems arise only from item nonresponse and response errors, not from interview nonresponse.

Second, we perform a "marginalist" analysis that supposes the entire design of the questionnaire has been set except for one item. The only decision is whether and how to ask about this item. Marginalist analysis enormously simplifies the decision problem. In practice, a survey planner must choose the entire structure of the questionnaire, and the choice made about one item may interact with choices made about others. We recognize this but, nevertheless, find it useful for exposition to focus on a single aspect of the global design problem, holding fixed the remainder of the questionnaire.

Third, we assume that the design chosen for the specific item in our marginalist analysis affects only the informativeness of that item. In practice, the choice of how to ask a specific item affects the length of the entire survey, which may influence respondents' willingness or ability to provide reliable responses to other items. We recognize this but, nevertheless, find it useful for exposition to suppose that the effect on other items is negligible.

Let $y$ denote the item under consideration. As indicated in the Introduction, the design options are as follows:

$A$: ask all respondents to report $y$.

$S$: ask only those respondents who respond positively to an opening question.

$N$: do not ask about $y$ at all.

The population parameter of interest is labeled $\tau[P(y)]$, where $P$ is the population distribution of $y$. For example, $\tau[P(y)]$ might be the population mean or median value of $y$.

3.2. *Measuring the cost, informativeness, and loss of the design options.* The design options differ in their costs and in their informativeness about $\tau[P(y)]$. Abstractly, let $c_k$ denote the cost of option $k$, let $d_k$ denote its informativeness, and let $L_k = L(c_k, d_k)$ be the loss that the survey planner associates with option $k$. We suppose that the planner wants to choose a design option that minimizes $L(c_k, d_k)$ over $k \in (A, S, N)$.

To operationalize this abstract optimization problem, a survey planner must decide how to measure loss, cost, and informativeness. Loss presumably increases with cost and decreases with informativeness. We will not be more specific about the form of the loss function here. We will, for simplicity, use a linear form in our applications.

Cost presumably increases with the fraction of respondents who are asked the item. In some settings, cost may be proportional to this fraction. Then $c_k = \gamma f_k$, where $\gamma > 0$ is the cost per respondent of data collection and $f_k$ is the fraction of respondents asked the item under option $k$. It is the case that $1 = f_A \geq f_S \geq f_N = 0$. Hence, $c_A = \gamma$, $c_S = \gamma f_S$, $c_N = 0$.

As indicated in the Introduction, we propose measurement of the informativeness of a design option by the size of the identification region obtained for the parameter of interest. In general, the size of an identification region depends on the specified parameter, the data produced by a design option, and the assumptions that the planner is willing to maintain. Sections 4 and 5 show how in some leading cases.

**4. Question design with nonresponse.** This section examines how nonresponse affects choice among the three design options. To focus attention on the inferential problem created by nonresponse, we assume that when sample members do respond, all answers are accurate. Section 4.1 considers identification of the parameter $\tau[P(y)]$. Section 4.2 shows how to use the findings to choose a design. Section 4.3 uses questions on future generosity of Social Security to illustrate.

4.1. *Identification with nonresponse.* It has been common in survey research to impute missing values and to use these imputations as if they are real data. Standard imputation methods presume that data are missing at random (MAR), conditional on specified observable covariates; see Little and Rubin (1987). If the maintained MAR assumptions are correct, then parameter $\tau[P(y)]$ is point-identified under both of design options $A$ and $S$. Option $S$ is less costly, so there is no reason to contemplate option $A$ from the perspective of identification. If option $A$ is used in practice, the reason must be to provide a larger sample of observations in order to improve statistical inference.

Identification becomes the dominant concern when, as is often the case, a survey planner has only a weak understanding of the distribution of missing data. We focus here on the worst-case setting, in which the planner knows nothing at all about the missing data. It is straightforward to determine the identification region for $\tau[P(y)]$ under design options $A$ and $S$. We draw on Manski [(2003), Chapter 1] to show how.

*Option A.* To formalize the identification problem created by nonresponse, let each member $j$ of a population $J$ have an outcome $y_j$ in a space $Y \equiv [0, s]$. Here $s$ can be finite or can equal $\infty$, in which case $Y$ is the nonnegative part of the extended real line. The assumption that $y$ is nonnegative is not crucial for our analysis, but it simplifies the exposition and notation.

The population is a probability space and $y: J \to Y$ is a random variable with distribution $P(y)$. Let a sampling process draw persons at random from $J$. However, not all realizations of $y$ are observable. Let the realization of a binary random variable $z_y^A$ indicate observability; $y$ is observable if $z_y^A = 1$ and not observable if $z_y^A = 0$. The superscript $A$ shows the dependence of observability of $y$ on design option $A$.

By the Law of Total Probability,

(1) $$P(y) = P(y|z_y^A = 1)P(z_y^A = 1) + P(y|z_y^A = 0)P(z_y^A = 0).$$

The sampling process reveals $P(y|z_y^A = 1)$ and $P(z_y^A)$, but it is uninformative regarding $P(y|z_y^A = 0)$. Hence, the sampling process partially identifies $P(y)$. In particular, it reveals that $P(y)$ lies in the identification region

(2) $$\mathrm{H}^A[P(y)] \equiv [P(y|z_y^A = 1)P(z_y^A = 1) + \psi P(z_y^A = 0), \psi \in \Psi_Y].$$

Here $\Psi_Y$ is the space of all probability distributions on $Y$ and the superscript $A$ on H shows the dependence of the identification region on the design option.

The identification region for a parameter of $P(y)$ follows immediately from $\mathrm{H}^A[P(y)]$. Consider inference on the parameter $\tau[P(y)]$. The identification region consists of all possible values of the parameter. Thus,

(3) $$\mathrm{H}^A\{\tau[P(y)]\} \equiv \{\tau(\eta), \eta \in \mathrm{H}^A[P(y)]\}.$$

Result (3) is simple but is too abstract to be useful as stated. Research on partial identification has sought to characterize $\mathrm{H}^A\{\tau[P(y)]\}$ for different parameters. Manski (1989) does this for means of bounded functions of $y$, Manski (1994) for quantiles, and Manski [(2003), Chapter 1] for all parameters that respect first-order stochastic dominance. Blundell et al. (2007) and Stoye (2005) characterize the identification regions for spread parameters such as the variance, interquartile range and the Gini coefficient.

The results for means of bounded functions are easy to derive and instructive, so we focus on these parameters here. To further simplify the exposition, we restrict attention to monotone functions. Let $\Re$ be the extended real line. Let $g(\cdot)$ be a monotone function that maps $Y$ into $\Re$ and that attains finite lower and upper bounds $g_0 \equiv \min_{y \in Y} g(y) = g(0)$ and $g_1 \equiv \max_{y \in Y} g(y)$. Without loss of generality, by a normalization, we set $g_0 = 0$ and $g_1 = 1$. The problem of interest is to infer $E[g(y)]$.

The Law of Iterated Expectations gives

(4) $$E[g(y)] = E[g(y)|z_y^A = 1]P(z_y^A = 1) + E[g(y)|z_y^A = 0]P(z_y^A = 0).$$

The sampling process reveals $E[g(y)|z_y^A = 1]$ and $P(z_y^A)$, but it is uninformative regarding $E[g(y)|z_y^A = 0]$, which can take any value in the interval $[0, 1]$. Hence, the identification region for $E[g(y)]$ is the closed interval

(5) $$\mathrm{H}^A\{E[g(y)]\} = [E[g(y)|z_y^A = 1]P(z_y^A = 1),$$
$$E[g(y)|z_y^A = 1]P(z_y^A = 1) + P(z_y^A = 0)].$$

$\mathrm{H}^A\{E[g(y)]\}$ is a proper subset of $[0, 1]$ whenever $P(z_y^A = 0)$ is less than one. The width of the region is $P(z_y^A = 0)$. Thus, the severity of the identification problem varies directly with the prevalence of missing data.

*Option S.* There are two sources of nonresponse under option $S$. First, a sample member may not respond to the opening question, in which case she is not asked about item $y$. Second, a sample member may respond to the opening question but not to the subsequent question about item $y$.

Let $x$ denote the item whose value is sought in the opening question. As in Illustrations 1 and 2, we suppose that $x$ is a broad item and that $y$ is a more specific one. For simplicity, we suppose here that $x \in \{0, 1\}$ and that $x = 0 \Longrightarrow y = 0$. A respondent is asked about $y$ only if she answers the opening question and reports $x = 1$. For example, consider Illustration 2 discussed in the Introduction. If a respondent does not have any limitation in ADLs ($x = 0$), then clearly the respondent does not have a limitation in bathing/showering ($y = 0$). Hence, the NLSOM asks about $y$ only when a respondent reports $x = 1$.

To formalize the identification problem, we need two response indicators, $z_x^S$ and $z_y^S$, the superscript $S$ showing the dependence of nonresponse on design option $S$. Let $z_x^S = 1$ if a respondent answers the opening question and let $z_x^S = 0$ otherwise. Let $z_y^S = 1$ if a respondent who is asked the follow-up question gives a response, with $z_y^S = 0$ otherwise. Hence, $z_y^S = 1 \Longrightarrow z_x^S = 1$. This and the Law of Iterated Expectations and the fact that $g(0) = 0$ give

$$\begin{aligned}
E[g(y)] &= E[g(y)|x=1]P(x=1) + E[g(y)|x=0]P(x=0) \\
&= E[g(y)|x=1, z_y^S=1]P(z_y^S=1, x=1) \\
&\quad + E[g(y)|x=1, z_x^S=1, z_y^S=0]P(z_x^S=1, z_y^S=0, x=1) \\
&\quad + E[g(y)|x=1, z_x^S=0]P(z_x^S=0, x=1).
\end{aligned}$$

The sampling process reveals $E[g(y)|x=1, z_y^S=1]$, $P(z_x^S=1, z_y^S=0, x=1)$, and $P(z_y^S=1) = P(z_y^S=1, x=1)$, where the last equality holds because $z_y^S = 1 \Longrightarrow x = 1$. The data are uninformative about $E[g(y)|x=1, z_x^S=1, z_y^S=0]$ and $E[g(y)|x=1, z_x^S=0]$, which can take any values in $[0, 1]$. The data are partially informative about $P(z_x^S=0, x=1)$, which can take any value in $[0, P(z_x^S=0)]$. It follows that the identification region for $E[g(y)]$ is the closed interval

$$\begin{aligned}
\text{H}^S\{E[g(y)]\} &= [E[g(y)|z_y^S=1]P(z_y^S=1), \\
&\quad E[g(y)|z_y^S=1]P(z_y^S=1) \\
&\quad + P(z_x^S=1, z_y^S=0, x=1) + P(z_x^S=0)].
\end{aligned}$$
(6)

Thus, the severity of the identification problem varies directly with the prevalence of nonresponse to the opening question and to the follow-up question in the subpopulation in which it is asked.

4.2. *Choosing a design.* Now consider choice among the three design options $(A, S, N)$. The widths of the identification regions for $E[g(y)]$ under these options are as follows:

$$d_A = P(z_y^A = 0), \qquad d_S = P(z_x^S = 1, z_y^S = 0, x = 1) + P(z_x^S = 0), \qquad d_N = 1.$$

For specificity, let the loss function have the linear form $L_k = \gamma f_k + d_k$. The first component measures survey cost and the second measures the informativeness of the design option. We set the coefficient on $d_k$ equal to one as a normalization of scale. The parameter $\gamma$ measures the importance that the survey planner gives to cost relative to informativeness. There is no universally "correct" value of this parameter. Its value is something that the survey planner must specify, depending on the survey context and the nature of item $y$.

It follows from the above and from the derivations of Section 4.1 that the losses associated with the three design options are as follows:

$$L_A = \gamma + P(z_y^A = 0),$$
$$L_S = \gamma P(z_x^S = 1, x = 1) + P(z_x^S = 1, z_y^S = 0, x = 1) + P(z_x^S = 0),$$
$$L_N = 1.$$

Thus, it is optimal to administer item $y$ to all sample members if

$$\gamma + P(z_y^A = 0) \leq \min\{1, \gamma P(z_x^S = 1, x = 1)$$
$$+ P(z_x^S = 1, z_y^S = 0, x = 1) + P(z_x^S = 0)\}.$$

Skip sequencing is optimal if

$$\gamma P(z_x^S = 1, x = 1) + P(z_x^S = 1, z_y^S = 0, x = 1) + P(z_x^S = 0)$$
$$\leq \min\{1, \gamma + P(z_y^A = 0)\}.$$

If neither of these inequalities hold, it is optimal not to ask the item at all.

Determination of the optimal design option requires knowledge of the response rates that would occur under options $A$ and $S$. This is where the body of survey research reviewed by Krosnick (1999) has a potentially important role to play. Through the use of randomized experiments embedded in surveys, researchers have developed considerable knowledge of the response rates that occur when various types of questions are posed to diverse populations. In many cases, this body of knowledge can be brought to bear to provide credible values for the response rates that determine loss under options $A$ and $S$.

When the literature does not provide credible values for these response rates, a survey planner may want to perform his own pretest, randomly assigning sample members to options $A$ and $S$. The size of the pretest sample only needs to be large enough to determine with reasonable confidence which design option is best. It does not need to be large enough to give precise estimates of the response rates.

4.3. *Questioning about expectations on the generosity of social security.* Consider the questions on expectations for the future generosity of the Social Security program cited in Illustration 1. The opening question was posed to 10,748 respondents to the 2006 HRS who currently receive social security benefits, and the follow-up was asked to the sub-sample of 9356 persons who answered the opening question and gave a response greater than zero. We assume here that the only data problem is nonresponse. The nonresponse rate to the opening question was 7.23%. The nonresponse rate to the follow-up question, for the subsample asked this question, was 2.27%. It is plausible that someone may not be willing to respond to the first question and yet be willing to respond to the second one. In particular, this would happen if a person does not want to speculate on what Congress will do but, nevertheless, is sure that if Congress does act, it would only change benefits for future retirees, not for those already in the system. The HRS use of skip sequencing prevents observation of $y$ in such cases.

To cast this application into the notation of the previous section, we let $x = 1$ if a respondent places a positive probability on Congress acting, with $x = 0$ otherwise. The rest of the notation is the same as above.

An early release of the HRS data provide these empirical values for the quantities that determine the identification region for $E[g(y)]$ and loss under design option $S$:

$$P(z_x^S = 1, z_y^S = 0, x = 1) = 0.0197,$$
$$P(z_x^S = 1, x = 1) = 0.8705,$$
$$P(z_y^S = 1) = 0.8508,$$
$$P(z_x^S = 0) = 0.0723,$$
$$E[g(y)|z_y^S = 1] = 0.4039,$$

where $g(y) \equiv \frac{y}{100}$. Hence, the identification region for $E[g(y)]$ under option $S$ is $\mathrm{H}^S\{E[g(y)]\} = [0.3436, 0.4356]$ and loss is $L_S = 0.8705\gamma + 0.0920$.

The HRS data do not reveal the quantities that determine the identification region for $E[g(y)]$ and loss under design option $A$. For this illustration, we conjecture that the mean response to item $y$ that would be obtained under option $A$ equals the mean response that is observed under option $S$. Thus, $E[g(y)|z_y^A = 1] = 0.4039$. We suppose further that the nonresponse probability would be $P(z_y^A = 0) = 0.08$. Then the identification region for $E[g(y)]$ under option $A$ is $\mathrm{H}^A\{E[g(y)]\} = [0.3716, 0.4516]$ and loss is $L_A = \gamma + 0.08$.

It follows from the above that it is optimal to administer item $y$ to all sample members if

$$\gamma \leq 0.0927.$$

Skip sequencing is optimal if

$$0.0927 \leq \gamma \leq 1.0431.$$

If neither of these inequalities hold, it is optimal not to ask the item at all.

**5. Question design with data errors.** This section examines how response errors affect choice among the three design options. To focus attention on the inferential problem created by such errors, we assume that all sample members respond to the questions posed. Section 5.1 considers identification. Section 5.2 shows how to use the findings to choose a design. Section 5.3 uses questions on limitations in ADLs to illustrate.

5.1. *Identification with response errors.* Section 4 showed that assumptions about the distribution of missing data are unnecessary for partially informative inference in the presence of nonresponse. In contrast, assumptions on the nature or prevalence of response errors are a prerequisite for inference. In cases where $y$ is discrete, it is natural to think of data errors as classification errors. We conceptualize response error here through a misclassification model previously used by Molinari (2003, 2008), and we draw on her findings. The Appendix discusses the mixture model of data errors, which yields equivalent results beginning from a different conceptualization of data errors.

The misclassification model is a simple formalism that does not have content per se. It becomes informative when it is combined with an assumed upper bound on the prevalence of data errors. When such a bound is available, Molinari (2003) showed that $E[g(y)]$ is partially identified under design option $A$. It is straightforward to show the same under option $S$. To simplify the exposition, we focus here on the particularly simple case where $y \in \{0,1\}$ and $g(y) \equiv y$. Corresponding results for general discrete $Y$ and any bounded function $g(\cdot): Y \to [0,1]$ may be obtained from the authors.

*Option A.* As in Section 4, let each member $j$ of a population $J$ have an outcome $y_j$ and let $P(y)$ be the population distribution of $y$. Let a sampling process draw persons at random from $J$. Let $\tilde{y}: J \to Y$ denote the responses that population members would give when queried about $y$. The researcher observes realizations of $\tilde{y}$, which can either equal or differ from the corresponding realizations of $y$. When $\tilde{y} \neq y$, data errors occur.

The misclassification model begins with the basic observation that, by the Law of Total Probability,

$$\begin{bmatrix} P(\tilde{y}^A = 1) \\ P(\tilde{y}^A = 0) \end{bmatrix} = \begin{bmatrix} P(\tilde{y}^A = 1|y=1) & P(\tilde{y}^A = 1|y=0) \\ P(\tilde{y}^A = 0|y=1) & P(\tilde{y}^A = 0|y=0) \end{bmatrix} \begin{bmatrix} P(y=1) \\ P(y=0) \end{bmatrix}.$$

The superscript $A$ shows that the response $\tilde{y}^A$ depends on design option $A$. The sampling process reveals only $P(\tilde{y}^A)$, which per se is uninformative about $P(y)$. The basic maintained assumption is a known nontrivial lower bound $1 - \lambda^A > 0$ on the probability that the realizations of $\tilde{y}^A$ and $y$ coincide, or, strengthening this assumption, a known nontrivial lower bound on the probability of correct report for each value that $y$ can take. Formally, these assumptions are as follows:

ASSUMPTION 1. $P(y = \tilde{y}^A) \geq 1 - \lambda^A > 0$.

ASSUMPTION 2. $P(\tilde{y}^A = k | y = k) \geq 1 - \lambda^A > 0, \ \forall \ k \in Y$.

Molinari (2003) shows that, under Assumption 1,

(7) $\quad \mathrm{H}^A[P(y=1)] = [0,1] \cap [P(\tilde{y}^A = 1) - \lambda^A, P(\tilde{y}^A = 1) + \lambda^A],$

while, under Assumption 2,

(8) $\quad \mathrm{H}^A[P(y=1)] = [0,1] \cap \left[ \dfrac{P(\tilde{y}^A = 1) - \lambda^A}{1 - \lambda^A}, \dfrac{P(\tilde{y}^A = 1)}{1 - \lambda^A} \right].$

Observe that these identification regions yield informative lower and upper bounds on $P(y=1)$ when $\lambda^A \leq P(\tilde{y}^A = 1) \leq 1 - \lambda^A$.

Results (7) and (8) were derived earlier by Horowitz and Manski (1995), using a different formalization of data errors. They studied partial identification of probability distributions under the mixture model of data errors used in studies of robust inference following Huber (1964). Their main assumption was the availability of an upper bound on the prevalence of data errors as defined in the mixture model, just as Huber assumed in his seminal research. See the Appendix for further discussion of the relationship between the mixture model and the misclassification model.

*Option $S$.* There are two sources of potential response error under option $S$. First, a sample member may respond with error to the opening question. Then she is erroneously not asked the follow up question if she gives a false negative answer, and she is erroneously asked the follow up question if she gives a false positive answer. Second, a sample member may (truthfully) respond affirmatively to the opening question and then respond with error to the follow up.

As in Section 4, we let $y$ denote the true value of the variable of interest and $x$ denote the true value of the variable elicited in the opening question. The error ridden versions of these variables are $\tilde{y}^S$ and $\tilde{x}^S$ respectively. As in Section 4, skip sequencing has certain logical implications when the opening question inquires broadly about a subject and the follow up inquires more specifically. These logical relations are $x = 0 \Longrightarrow y = 0$ and $\tilde{x}^S = 0 \Longrightarrow \tilde{y}^S = 0$.

The misclassification model begins with the observation that, by the Law of Total Probability,

$$P(\tilde{x}^S = i, \tilde{y}^S = k)$$
$$= \sum_{l=0,1} \sum_{m=0,1} P(\tilde{x}^S = i, \tilde{y}^S = k | x = l, y = m) P(x = l, y = m),$$

$$i, k \in \{0, 1\}.$$

The sampling process reveals only the quantities $P(\tilde{x}^S = i, \tilde{y}^S = k)$ on the left-hand side of these equations, with the logic of skip sequencing implying that $P(\tilde{x}^S = 1, \tilde{y}^S = 1) = P(\tilde{y}^S = 1)$, $P(\tilde{x}^S = 0, \tilde{y}^S = 0) = P(\tilde{x}^S = 0)$ and $P(\tilde{x}^S = 0, \tilde{y}^S = 1) = 0$. The logic of skip sequencing also implies that $P(x = 1, y = 1) = P(y = 1)$, $P(x = 0, y = 0) = P(x = 0)$ and $P(x = 0, y = 1) = 0$.

The observable quantities and logical restrictions per se are uninformative about $P(y)$, but they become informative when combined with these extensions of Assumptions 1 and 2:

ASSUMPTION 3.   $P(x = \tilde{x}^S, y = \tilde{y}^S) \geq 1 - \lambda^S > 0$.

ASSUMPTION 4.   $P(\tilde{x}^S = i, \tilde{y}^S = k | x = i, y = k) \geq 1 - \lambda^S > 0$, $i, k \in \{0, 1\}$, $k \leq i$.

Extension of the argument of Molinari (2003) shows that, under Assumptions 3,

(9) $\quad \mathrm{H}^S[P(y=1)] = [0,1] \cap [P(\tilde{y}^S = 1) - \lambda^S, P(\tilde{y}^S = 1) + \lambda^S],$

while, under Assumption 4,

(10) $\quad \mathrm{H}^S[P(y=1)] = [0,1] \cap \left[ \dfrac{P(\tilde{y}^S = 1) - \lambda^S}{1 - \lambda^S}, \dfrac{P(\tilde{y}^S = 1)}{1 - \lambda^S} \right].$

These identification regions yield informative lower and upper bounds on $P(y = 1)$ when $\lambda^S \leq P(\tilde{y}^S = 1) \leq 1 - \lambda^S$.

Whereas Assumptions 1 and 2 only concerned the coincidence of the true and reported values of $y$, Assumptions 3 and 4 concern the joint coincidence of the true and reported values of $(x, y)$. Hence, it is reasonable to think that a survey planner will ordinarily specify a higher lower bound in the first case than the second; that is, $1 - \lambda^A > 1 - \lambda^S$.

5.2. *Choosing a design.* Now consider choice among the three design options. The width of the identification region for $P(y=1)$ under option $N$ remains $d_N = 1$, and therefore, the loss associated with this option is $L_N = 1$.

For simplicity, we focus here on the case when the identification regions under Options $A$ and $S$ yield informative lower and upper bounds; that is, $\lambda^k \leq P(\tilde{y}^k = 1) \leq 1 - \lambda^k, k \in (A, S)$. Table 1 contains the results for other cases.

Under Assumptions 1 and 3, the widths of the identification regions for $P(y=1)$, under design options $A$ and $S$, are $d_k = 2\lambda^k$, $k \in (A, S)$. Therefore, the losses associated with these two design options are

$$L_A = \gamma + 2\lambda^A, \qquad L_S = \gamma P(\tilde{x}^S = 1) + 2\lambda^S.$$

Thus, it is optimal to ask about item $y$ to all sample members if

$$\gamma + 2\lambda^A \leq \min\{1, \gamma P(\tilde{x}^S = 1) + 2\lambda^S\}.$$

Skip sequencing is optimal if

$$\gamma P(\tilde{x}^S = 1) + 2\lambda^S \leq \min\{1, \gamma + 2\lambda^A\}.$$

If neither of these inequalities hold, it is optimal not to ask the item at all.

Under Assumptions 2 and 4, the widths of the identification regions for $P(y=1)$ are $d_k = \frac{\lambda^k}{1-\lambda^k}$, $k \in (A, S)$. Therefore, the losses are

$$L_A = \gamma + \frac{\lambda^A}{1 - \lambda^A}, \qquad L_S = \gamma P(\tilde{x}^S = 1) + \frac{\lambda^S}{1 - \lambda^S}.$$

Thus, it is optimal to ask about item $y$ to all sample members if

$$\gamma + \frac{\lambda^A}{1 - \lambda^A} \leq \min\left\{1, \gamma P(\tilde{x}^S = 1) + \frac{\lambda^S}{1 - \lambda^S}\right\}.$$

Table 1
*Value of $L_k$ depending on the relationship between $\lambda^k$ and $P(\tilde{y}^k = 1)$, $k \in (A, S)$*

| | Assumptions 1 and 3 | Assumptions 2 and 4 |
|---|---|---|
| $1 - \lambda^A \leq P(\tilde{y}^A = 1) \leq \lambda^A$ | $L_A = \gamma + 1$ | $L_A = \gamma + 1$ |
| $P(\tilde{y}^A = 1) \leq \min\{\lambda^A, 1 - \lambda^A\}$ | $L_A = \gamma + P(\tilde{y}^A = 1) + \lambda^A$ | $L_A = \gamma + \frac{P(\tilde{y}^A = 1)}{1 - \lambda^A}$ |
| $\lambda^A \leq P(\tilde{y}^A = 1) \leq 1 - \lambda^A$ | $L_A = \gamma + 2\lambda^A$ | $L_A = \gamma + \frac{\lambda^A}{1 - \lambda^A}$ |
| $P(\tilde{y}^A = 1) \geq \max\{\lambda^A, 1 - \lambda^A\}$ | $L_A = \gamma + 1 - P(\tilde{y}^A = 1) + \lambda^A$ | $L_A = \gamma + \frac{1 - P(\tilde{y}^A = 1)}{1 - \lambda^A}$ |
| $1 - \lambda^S \leq P(\tilde{y}^S = 1) \leq \lambda^S$ | $L_S = \gamma \delta_x^S + 1$ | $L_S = \gamma \delta_x^S + 1$ |
| $P(\tilde{y}^S = 1) \leq \min\{\lambda^S, 1 - \lambda^S\}$ | $L_S = \gamma \delta_x^S + P(\tilde{y}^S = 1) + \lambda^S$ | $L_S = \gamma \delta_x^S + \frac{P(\tilde{y}^S = 1)}{1 - \lambda^S}$ |
| $\lambda^S \leq P(\tilde{y}^S = 1) \leq 1 - \lambda^S$ | $L_S = \gamma \delta_x^S + 2\lambda^S$ | $L_S = \gamma \delta_x^S + \frac{\lambda^S}{1 - \lambda^S}$ |
| $P(\tilde{y}^S = 1) \geq \max\{\lambda^S, 1 - \lambda^S\}$ | $L_S = \gamma \delta_x^S + 1 - P(\tilde{y}^S = 1) + \lambda^S$ | $L_S = \gamma \delta_x^S + \frac{1 - P(\tilde{y}^S = 1)}{1 - \lambda^S}$ |

NOTE. $\delta_x^S \equiv P(\tilde{x}^S = 1)$.

Skip sequencing is optimal if

$$\gamma P(\tilde{x}^S = 1) + \frac{\lambda^S}{1-\lambda^S} \leq \min\bigg\{1, \gamma + \frac{\lambda^A}{1-\lambda^A}\bigg\}.$$

If neither of these inequalities hold, it is optimal not to ask the item at all.

Determination of the optimal design option requires information on the nature and prevalence of response errors under options $A$ and $S$. There have been occasional validation and reliability studies documenting the extent of measurement error in survey items; see, for example, Groves (1989) and Bound, Brown and Mathiowetz (2001). When the literature does not provide credible upper bounds for the probability of data errors, a survey planner may want to perform his own pretest, randomly assigning sample members to options $A$ and $S$, and then obtain corresponding validation or reliability data. As in Section 4, the size of the pretest sample only needs to be large enough to determine with reasonable confidence which design option is best. It does not need to be large enough to give precise estimates of the upper bounds on the probabilities of data errors.

5.3. *Questioning about limitations in ADLs.* Consider the questions on limitations in ADLs cited in Illustration 2. The opening question was posed to 2092 respondents to the 1990 NLSOM, of whom 92.45% were self respondents and 7.55% were proxy respondents. The follow-ups were asked to the 192 persons who responded to the opening question and gave an affirmative answer. We focus here on the first follow-up ADL question: "Now I would like to be more specific. Because of a health or physical problem, do you receive help from another person in bathing or showering?" The nonresponse rate to the opening question was 0.62%. The nonresponse rate to the follow-up question, for the subsample asked this question, was 0.52%. Given these minimal nonresponse rates, we abstract from nonresponse here and concentrate our attention on response error.

To keep this illustration simple, we suppose here that the question on bathing or showering is the only follow up to the NLSOM opening question on limitations in ADLs. A more realistic analysis would jointly consider the six follow up questions that actually appear in the survey. This is a straightforward extension of our analysis if one maintains the "marginalist" assumption that the design chosen for the set of ADL items does not affect data quality elsewhere in the survey. We think this assumption reasonable, because the NLSOM contains only six easily understood questions on limitations in specific ADLs. Item nonresponse to these questions is minimal. Item nonresponse also was minimal when similar questions were asked in the AHEAD survey, described below, which does not use skip sequencing.

We caution that there are circumstances in which skip sequencing avoids having to ask some respondents a long, laborious sequence of irrelevant questions. When this is the case one may, as noted in Section 3.1, think that the skip sequencing decision may materially affect respondents' willingness or ability to provide reliable responses throughout the survey. When respondent burden is a potential concern, one may find it necessary to move away from simple marginalist analysis of the type we perform and instead treat the design of the entire questionnaire as a complex joint decision problem.

For this illustration, we take the parameter of interest to be the cross-sectional probability $P(y = 1)$ that an individual in the population represented by the NLSOM needs help in bathing/showering. This is one of several parameters of potential interest when studying limitations in ADL. Connor et al. (2006) emphasize the importance of longitudinal measurement of the duration of disability and of transitions in and out of disability. Concern with these matters might lead one to be interested in $P[y(t) - y(t-k)]$ or $P[y(t)|y(t-k)]$, where $y(t)$ and $y(t-k)$ measure limitations in ADLs at two interviews spaced $k$ years apart. It would be of interest to characterize the identification regions for these transition parameters under alternative questionnaire designs.

Consider $P(y = 1)$. The reported probability is $P(\tilde{y}^S = 1) = 0.073$. To apply the misclassification model, we need to set values for the upper bounds $\lambda^A$ and $\lambda^S$ on the probability of occurrence of data errors under options $A$ and $S$. We are not aware of validation studies placing upper bounds on the probability of data errors in self reports of limitations in ADLs for populations similar to the one surveyed by the NLSOM, under design option $S$. However, there have been studies that compare self reports and proxy reports, as well as some that assess the time series consistency of self reports across interviews. Most of this work analyzes surveys in which the questionnaire uses design option $A$. See, for example, Rubenstein et al. (1984), Mathiowetz and Groves (1985), Moore (1988), Mathiowetz and Lair (1994), Rodgers and Miller (1997), Mathiowetz and Wunderlich (2000) and Miller and DeMaio (2006). In particular, Rubenstein et al. (1984) and Miller and DeMaio (2006) report the results of reliability studies providing information on the prevalence of data errors.

Rubenstein et al. (1984) analyze two samples of individuals, one providing data on hospitalized elderly persons and the other on nursing home residents. They compare the reports of limitations in ADLs and additional daily activities (such as telephoning, shopping, handling finances, cooking, etc.) of the institutionalized elderlies and of a "community proxy" (a spouse, child, or close friend) with those of a nurse proxy. If one assumes that the report of the nurse proxy is always correct, one can conclude from this study that the probability of a data error is bounded above by 0.36. Miller and DeMaio (2006) analyze data on limitations in bathing/showering collected in the 2006 administration of the American Community Survey Content Test. Reliability estimates based on reinterviews suggest a probability of data errors of at most 0.17.

The sampling frame and questionnaire design of the NLSOM differ from the ones analyzed in these reliability studies. Hence, their findings can only be suggestive for our purposes. In what follows we use the bounds in Assumption 5 below. Table 2 collects the results obtained using different values of $\lambda^A$ and $\lambda^S$, which encompass the upper bounds on probabilities of data errors reported by Rubenstein et al. (1984) and Miller and DeMaio (2006).

ASSUMPTION 5.  $\lambda^A = 0.15,\ \lambda^S = 0.25$.

The identification regions for $P(y=1)$ under design options $A$ and $S$ are given in Table 1. [The forms given in Section 5.2 do not apply here because the inequalities $\lambda^k \leq P(\tilde{y}^k = 1) \leq 1 - \lambda^k, k \in (A, S)$ do not hold in this application.] Using $\lambda^S = 0.25$ as the upper bound on data errors under design option $S$, the identification region for $P(y=1)$ is $\mathrm{H}^S[P(y=1)] = [0, 0.3230]$ under Assumption 3 and $\mathrm{H}^S[P(y=1)] = [0, 0.0973]$ under Assumption 4. The data reveal that $P(\tilde{x}^S = 1) = 0.092$. Hence, loss is $L_S = 0.092\gamma + 0.3230$ under Assumption 3, and $L_S = 0.092\gamma + 0.0973$ under Assumption 4.

The NLSOM data do not reveal the quantity $P(\tilde{y}^A = 1)$ needed to determine the identification region for $P(y=1)$ under design option $A$. For this illustration, we conjecture that the rate of reported limitations in bathing/showering that would be obtained under option $A$ equals the rate that is observed under option $S$. Thus, $P(\tilde{y}^A = 1) = 0.073$. Using $\lambda^A = 0.15$ as the upper bound on data errors under option $A$, the identification region for $P(y=1)$ is $\mathrm{H}^A[P(y=1)] = [0, 0.2230]$ under Assumption 1 and $\mathrm{H}^A[P(y=1)] = [0, 0.0859]$ under Assumption 2. Hence, loss is $L_A = \gamma + 0.2230$ under Assumption 1 and $L_A = \gamma + 0.0859$ under Assumption 2.

It follows that it is optimal to ask all sample member about item $y$ if

$$\gamma + 0.2230 \leq \min\{1, 0.092\gamma + 0.3230\} \quad \Longleftrightarrow \quad \gamma \leq 0.1101$$

under Assumptions 1 and 3,

$$\gamma + 0.0859 \leq \min\{1, 0.092\gamma + 0.0973\} \quad \Longleftrightarrow \quad \gamma \leq 0.0126$$

under Assumptions 2 and 4.

Skip sequencing is optimal if

$$0.092\gamma + 0.3230 \leq \min\{1, \gamma + 0.2230\} \Longleftrightarrow 0.1101 \leq \gamma \leq 7.3587$$

under Assumptions 1 and 3,

$$0.092\gamma + 0.0973 \leq \min\{1, \gamma + 0.0859\} \Longleftrightarrow 0.0126 \leq \gamma \leq 9.8116$$

under Assumptions 2 and 4.

Otherwise, it is optimal not to ask the item at all.

TABLE 2
Values of $\gamma$ that determine the choice of a certain design option, depending on $(\lambda^A, \lambda^S)$

| $\lambda^A$ | $\lambda^S$ | Assumptions 1 and 3 | | Assumptions 2 and 4 | |
| --- | --- | --- | --- | --- | --- |
| | | Option $A$ is chosen | Option $S$ is chosen | Option $A$ is chosen | Option $S$ is chosen |
| 0.100 | 0.100 | Never | $0.000 \leq \gamma \leq 8.989$ | Never | $0.000 \leq \gamma \leq 9.988$ |
| | 0.125 | $\gamma \leq 0.027$ | $0.027 \leq \gamma \leq 8.717$ | $\gamma \leq 0.003$ | $0.003 \leq \gamma \leq 9.963$ |
| | 0.170 | $\gamma \leq 0.077$ | $0.077 \leq \gamma \leq 8.228$ | $\gamma \leq 0.007$ | $0.007 \leq \gamma \leq 9.914$ |
| | 0.200 | $\gamma \leq 0.110$ | $0.110 \leq \gamma \leq 7.902$ | $\gamma \leq 0.011$ | $0.011 \leq \gamma \leq 9.878$ |
| | 0.360 | $\gamma \leq 0.286$ | $0.286 \leq \gamma \leq 6.163$ | $\gamma \leq 0.036$ | $0.036 \leq \gamma \leq 9.630$ |
| | 0.400 | $\gamma \leq 0.330$ | $0.330 \leq \gamma \leq 5.728$ | $\gamma \leq 0.045$ | $0.045 \leq \gamma \leq 9.547$ |
| 0.125 | 0.125 | Never | $0.000 \leq \gamma \leq 8.717$ | Never | $0.000 \leq \gamma \leq 9.963$ |
| | 0.170 | $\gamma \leq 0.050$ | $0.050 \leq \gamma \leq 8.228$ | $\gamma \leq 0.005$ | $0.005 \leq \gamma \leq 9.914$ |
| | 0.200 | $\gamma \leq 0.083$ | $0.083 \leq \gamma \leq 7.902$ | $\gamma \leq 0.008$ | $0.008 \leq \gamma \leq 9.878$ |
| | 0.360 | $\gamma \leq 0.259$ | $0.259 \leq \gamma \leq 6.163$ | $\gamma \leq 0.034$ | $0.034 \leq \gamma \leq 9.630$ |
| | 0.400 | $\gamma \leq 0.303$ | $0.303 \leq \gamma \leq 5.728$ | $\gamma \leq 0.042$ | $0.042 \leq \gamma \leq 9.547$ |
| 0.170 | 0.170 | Never | $0.000 \leq \gamma \leq 8.228$ | Never | $0.000 \leq \gamma \leq 9.914$ |
| | 0.200 | $\gamma \leq 0.033$ | $0.033 \leq \gamma \leq 7.902$ | $\gamma \leq 0.004$ | $0.004 \leq \gamma \leq 9.878$ |
| | 0.360 | $\gamma \leq 0.209$ | $0.209 \leq \gamma \leq 6.163$ | $\gamma \leq 0.029$ | $0.029 \leq \gamma \leq 9.630$ |
| | 0.400 | $\gamma \leq 0.253$ | $0.253 \leq \gamma \leq 5.728$ | $\gamma \leq 0.037$ | $0.037 \leq \gamma \leq 9.547$ |
| 0.200 | 0.200 | Never | $0.000 \leq \gamma \leq 7.902$ | Never | $0.000 \leq \gamma \leq 9.878$ |
| | 0.360 | $\gamma \leq 0.176$ | $0.176 \leq \gamma \leq 6.163$ | $\gamma \leq 0.025$ | $0.025 \leq \gamma \leq 9.630$ |
| | 0.400 | $\gamma \leq 0.220$ | $0.220 \leq \gamma \leq 5.728$ | $\gamma \leq 0.033$ | $0.033 \leq \gamma \leq 9.547$ |
| 0.360 | 0.360 | Never | $0.000 \leq \gamma \leq 6.163$ | Never | $0.000 \leq \gamma \leq 9.630$ |
| | 0.400 | $\gamma \leq 0.044$ | $0.044 \leq \gamma \leq 5.728$ | $\gamma \leq 0.008$ | $0.008 \leq \gamma \leq 9.547$ |
| 0.400 | 0.400 | Never | $0.000 \leq \gamma \leq 5.728$ | Never | $0.000 \leq \gamma \leq 9.547$ |

We conclude this section by calling attention to the fact that the 1993 wave of the Assets and Health Dynamics Among the Oldest Old (AHEAD) survey targeted a population similar in age to the NLSOM. The AHEAD survey also asked respondents about their limitations in ADLs, but it used neither design option $A$ or $S$. Instead, AHEAD omitted the opening broad question of the NLSOM and immediately posed a series of specific questions to all respondents. The fraction of AHEAD respondents who reported limitations in bathing/showering was 0.085, a value close to that elicited in the NLSOM. To compare the AHEAD and NLSOM designs would require generalization of the decision problem that we set up in Section 3. In particular, we would need to take into account the loss of information on limitations in ADLs that may potentially occur in AHEAD by dropping the opening question.

**6. Conclusion.** Survey planners have long had to cope with the tension between the desire to reduce the costs and increase the informativeness of surveys. However, they have not studied questionnaire design as a formal decision problem in which one uses an explicit loss function to quantify the trade-off between cost and informativeness. Groves (1987) called attention to this in an article in *Public Opinion Quarterly* (POQ), writing (page S167):

> "The inextricable link between costs and errors rarely is formally acknowledged in methods articles in POQ, or in any other scholarly journal for that matter. That state of affairs has two detrimental effects: (1) methodologists invent methods to reduce an error, but fail to measure the cost impact of the new idea, and (2) practitioners reject new ideas until it becomes clear that they result in reduced costs. Given the link between errors and costs, many new ideas require spending money to reduce an error."

Groves went on to contrast the situation in questionnaire design with that in survey sampling, which has long used formal models of cost and sampling error to analyze the problem of choosing sample size. See also Spencer (1980, 1985, 1994), who has argued broadly for benefit–cost analysis of programs of data collection, with particular attention to the U.S. Census.

This paper has formally analyzed skip sequencing as a decision problem in questionnaire design. We have intentionally kept the exposition simple in order to highlight the basic trade-off between cost and informativeness in choosing a design option. Survey researchers and statisticians with traditional training may be least familiar with our measurement of informativeness by the size of the identification region for a population parameter of interest. Although identification is the central problem generated by nonresponse and response errors, the research literatures in survey research and statistics contain remarkably little formal analysis of identification. We think that the illustrative cases considered in Sections 4 and 5 give a constructive sense of how to proceed, without getting bogged down in mathematical detail.

While identification is the dominant issue in assessing data quality in large surveys, sampling error can also be a significant concern in smaller surveys. A straightforward extension of our work to smaller surveys is to measure informativeness through a confidence interval for the partially identified parameter of interest. The literature on partial identification has recently spawned many approaches to the construction of asymptotically valid confidence intervals. See, for example, Imbens and Manski (2004), Chernozhukov, Hong and Tamer (2007) and Beresteanu and Molinari (2008). Another approach, with a firmer decision-theoretic foundation, would be to address the questionnaire design problem from the perspective of Wald (1950).

## APPENDIX: MIXTURE MODEL AND MISCLASSIFICATION MODEL

The mixture model of robust statistics introduces latent variables $e \in Y$ and $w \in \{0,1\}$, and views the reported values $\tilde{y}$ as generated by the mixture $\tilde{y} = wy + (1-w)e$. The unobservable binary variable $w$ denotes whether $y$ or $e$ is observed. Realizations of $\tilde{y}$ with $w=1$ are said to be error free and those with $w=0$ are said to be data errors. By the Law of Total Probability, the relationship between the observable distribution $P(\tilde{y})$ and the unobservable distribution $P(y)$ is

(11) $$P(\tilde{y}) = P(y|w=1)P(w=1) + P(e|w=0)P(w=0),$$

(12) $$P(y) = P(y|w=1)P(w=1) + P(y|w=0)P(w=0).$$

The mixture model per se is a formalism without content. It becomes informative when accompanied by assumption of an upper bound on the occurrence of data errors, as follows:

ASSUMPTION A.1. $P(w=0) \leq \lambda < 1$.

It is sometimes also assumed that the occurrence of errors is statistically independent of the value of $y$. That is,

ASSUMPTION A.2. $y \perp w$.

Horowitz and Manski (1995) studied the implications of the mixture model for partial identification of probability distributions; see also Manski (2003), Chapter 4. They derived the identification region for $P(y)$ and for parameters of this distribution that respect stochastic dominance, under Assumption A.1 alone and under Assumptions A.1 and A.2. They refer to the first case as "corrupted sampling," and to the second as "contaminated sampling."

The relationship between the mixture model and the misclassification model can be easily established starting from equation (11). Observe that

(13) $$P(\tilde{y}=j|y=k)$$
$$= \begin{cases} P(w=1|y=k) + P(e=k|y=k,w=0)P(w=0|y=k), & \text{if } j=k, \\ P(e=j|y=k,w=0)P(w=0|y=k), & \text{if } j \neq k. \end{cases}$$

Hence, assumptions on $P(w|y)$ translate immediately into assumptions for the misclassification model. Molinari (2003) shows that if the distribution of $e$ is unrestricted, the mixture model with Assumptions A.1 and A.2 is equivalent to the misclassification model with an assumption specifying a common lower bound on the probabilities of correct report, $P(\tilde{y}=k|y=k)$, $k \in Y$. The mixture model with Assumption A.1 alone is equivalent to the misclassification model with an assumption specifying a lower bound on the probability that $\tilde{y}$ and $y$ coincide, $P(\tilde{y}=y)$.

**Acknowledgments.** We thank four anonymous reviewers and the Editor for comments.

Department of Economics
and
Institute for Policy Research
Northwestern University
2001 Sheridan Road
Evanston, Illinois 60208-2600
USA
E-mail: cfmanski@northwestern.edu

Department of Economics
Cornell University
Uris Hall
Ithaca, New York 14853-7601
USA
E-mail: fm72@cornell.edu